\documentclass{elsart}

\newif\ifpdf
    \ifx\pdfoutput\undefined
    \pdffalse    
\else
    \pdfoutput=1 
    \pdftrue
\fi \ifpdf
    \usepackage[pdftex]{graphicx}
\else
    \usepackage{graphicx}
\fi
\usepackage{amsfonts}
\usepackage{amssymb}
\usepackage{amsmath}
\usepackage[all]{xy}
\begin{document}
\ifpdf
    \DeclareGraphicsExtensions{.pdf}
\else
    \DeclareGraphicsExtensions{.eps}
\fi

\begin{frontmatter}

\title{Un-particle Effective Action}

\author{Patricio Gaete\thanksref{cile}}
\thanks[cile]{e-mail address: patricio.gaete@usm.cl}
\address{Departamento de F\'{\i}sica, Universidad T\'ecnica
Federico Santa Mar\'{\i}a, Valpara\'{\i}so, Chile}

\author{Euro Spallucci\thanksref{infn}}
\thanks[infn]{e-mail address: spallucci@ts.infn.it }
\address{Dipartimento di Fisica Teorica, Universit\`a di Trieste
and INFN, Sezione di Trieste, Italy}

\begin{abstract}
We study un-particle dynamics in the framework of standard quantum
field theory. We obtain the Feynman propagator by supplementing
standard quantum field theory definitions with integration over the
mass spectrum. Then we use this information to construct effective
actions for scalar, gauge vector and gravitational un-particles.
\end{abstract}
\end{frontmatter}

\section{Introduction}
In a seminal paper \cite{Banks:1981nn} Banks and Zaks investigated
the unusual properties of matter with non-trivial scale invariance
in the infra-red regime. Contrary to the intuitive notion of scale
invariance as a property of massless particles only, this new kind
of stuff has no definite mass at all. For this reason, this
presently unknown, scale invariant, sector of the elementary
particle spectrum has been dubbed as the \textit{un-particle} sector
\cite{Georgi:2007ek,Georgi:2007si} . \\
In a  recent couple of papers \cite{Georgi:2007ek,Georgi:2007si}
it has been proposed that below
some critical energy scale $\Lambda_U$ the standard model particles
can interact with un-particles. The forthcoming start of LHC activity
has focused the interest of the high energy physics
community on possible experimental signature of un-particles events
at the $TeV$ energy scale
\cite{Strassler:2008bv,Liao:2007ic,Liao:2007fv,Liao:2007bx,Rizzo:2007xr}
\cite{Cheung:2007ap,Bander:2007nd,Cheung:2007jb,Cheung:2007zza}
\cite{Kikuchi:2007qd,Kikuchi:2008pr}.
Astroparticle and cosmological unparticle effects have been considered as well
\cite{Kikuchi:2007az,Lewis:2007ss,McDonald:2007bt,Davoudiasl:2007jr},
\cite{Das:2007nu,Alberghi:2007vc,Hannestad:2007ys,Freitas:2007ip,Chen:2007qc}.\\
On the  theoretical side,  interesting connections
between un-particles and non-standard Kaluza-Klein dynamics in
extra-dimensions, and AdS/CFT duality, have been pointed out in
\cite{Stephanov:2007ry,Kazakov:2007fn,Kazakov:2007su,Lee:2007xd}.
Still in the background
is the dynamics of unparticle sector itself.\\
In this letter we would like to consider un-particle dynamics in the
familiar language of standard (effective) quantum field theory in
four dimensions. This piece of information can be useful in view of
studying gravitational effects \cite{Goldberg:2007tt,Mureika:2007nc}
beyond the weak field approximation.\\
In Sect.2 we use path integral techniques to recover Euclidean
propagator, static potential and effective action for scalar and
Abelian gauge vector un-particle fields. In Sect.3 we provide an
alternative evaluation of the static potential through the
expectation value of the Hamiltonian between source physical states.
In Sect.4 we extend the results of Sect.2 to the non-Abelian case
and to gravity; the final result is the effective
action for the un-graviton.\\

\section{Effective Action}

We start from the analogy between un-particles and continuous mass
spectrum objects \cite{Krasnikov:2007fs} and implement this relation
in the Euclidean functional integral.

\begin{equation}
Z_U\left(\, J\,\right)\equiv \frac{A_{d_U}}{2\,
\left(\,\Lambda^2_U\,\right)^{d_U-1}}\,
\int_0^\infty dm^2
\, \left(\, m^2\,\right)^{d_U-2}\, Z\left(\, J\,\right)
\label{uz}
\end{equation}

\begin{equation}
A_{d_U}\equiv \frac{16\,\pi^{5/2}}{\left(\, 2\pi\,\right)^{2d_U}}\,
\frac{\Gamma\left(\, d_U +1/2\,\right)}{\Gamma\left(\, d_U -1\,\right)\,
 \Gamma\left(\, 2d_U \,\right)} \label{uz2}
\end{equation}

where $d_U$ is a non integral scale dimension of the un-particle
field. This parameter is a distinctive feature of un-particle
physics which looks like a sort of \textit{fractal} extension of
ordinary particle physics. As fractal geometry allows to approximate
non-integer dimension surfaces, un-particle physics provides an
effective description of the quantum dynamics of a non-integer
number of non-separable quanta. As a reasonable working hypothesis,
it is generally assumed $1 < d_U\le 2$, even if, on a general
ground, higher values of $d_U$ cannot be dismissed.
\\
$Z\left(\, J\,\right)$ is the generating functional for a
``particle'' quantum field theory. Formula (\ref{uz}) establishes
the connection between particle and un-particle quantum field
theory. The corresponding Green functions are related in the same
way

\begin{eqnarray}
G_U\left(\, x-y\,\right)&&=\left[\,\frac{\delta^2 Z_U\left(\, J\,\right)}{
\delta\, J\left(\, x\,\right) \,\delta J \left(\, y\,\right)\, }\,\right]_{J=0}
\nonumber\\
&&=\frac{A_{d_U}}{2\,
\left(\,\Lambda^2_U\,\right)^{d_U-1}}\,
\int_0^\infty dm^2\, \left(\, m^2\,\right)^{d_U-2}\,
G\left(\, x-y\ ; m^2\,\right)\label{gu}
\end{eqnarray}

As a check of our approach let us compute the static potential generated
by the exchange of scalar un-particles.  Un-particles give rise to
long-range forces, thus it is interesting to determine how they modify
gravitational and Coulombic interactions \cite{Deshpande:2007mf}.\\
For $Z_\phi\left(\, J\,\right)$ we consider

\begin{equation}
 Z_\phi\left(\, J\,\right)=\int \left[\, D\phi\,\right]\,\exp\left[\, -\int
 d^4x\, \left(\, \frac{1}{2}\partial_\mu\phi\,\partial^\mu\phi +m^2\,\phi^2
 + J\,\phi\,\right)\,\right]
 \end{equation}

where the appropriate normalization factor is understood in the
functional integration measure. The gaussian integration leads to
the known result

\begin{equation}
Z_\phi\left(\, J\,\right)=\exp\left[-\frac{1}{2}\int d^4x\,  \int d^4y
J\left(\, x\,\right)\frac{1}{-\partial^2 +m^2}\,\delta\left(\, x-y\,\right)
\, J\left(\, y\,\right)\,\right]\label{zj}
\end{equation}

On the other hand, $Z_\phi\left(\, J\,\right)$ represents the average
value of the exponential interaction energy between the field and the source:

\begin{equation}
Z_\phi\left(\, J\,\right)=\big\langle \, \exp\left[-\int d^4x\, J\,\phi\,\right]\,
\big\rangle
\end{equation}

For ``infinitely heavy'' sources interaction energy reduces to
average of  the static potential defined as

\begin{equation}
Z_\phi\left[\, V\left(\,\vec{x}\,\right)\,\right]
=\big\langle\, \exp\left[-\frac{1}{8\pi}\,
\int_0^T dt\,V\left[\,\vec{x}\,\right)\, \right]\,\big\rangle
\label{pstat}
\end{equation}

Comparison between  (\ref{zj}) and (\ref{pstat}) gives

\begin{equation}
V_U\left(\, \vec{x}\,\right)= 4\pi\, \int d^3y\, G_U\left(\, \vec{x}-\vec{y}
\,\right)\,j\left(\,\vec{y}\,\right)\label{pot}
\end{equation}

For a (static) point-like source, located in the origin,
$j\left(\,\vec{y}\,\right)= \kappa\, \delta\left(\, \vec{y} \, \right)$,
and Eq.(\ref{pot}) gives

\begin{equation}
V_U\left(\, \vec{x}\,\right)= 4\pi\,\kappa \, G_U\left(\, \vec{x}\,\right)\,
=4\pi\,\kappa \,\int \frac{d^3p}{\left(\, 2\pi\,\right)^3}\,
e^{i\vec{p}\cdot\vec{x}}\, G_U\left(\, \vec{p}^{\,2}\,\right)
\label{pot2}
\end{equation}

>From the definition (\ref{gu}) we get

\begin{eqnarray}
 G_U\left(\, p^{\,2}\,\right)&&=-\frac{A_{d_U}}{2\,
\left(\,\Lambda^2_U\,\right)^{d_U-1}}\,
\int_0^\infty dm^2
\, \left(\, m^2\,\right)^{d_U-2}\,\int_0^\infty ds\, e^{-s\left(\, p^2
+m^2\,\right)}\nonumber\\
&&= -\frac{A_{d_U}}{2\pi\, \left(\,\Lambda^2_U\,\right)^{d_U-1}}
\,\Gamma\left(\, d_U-1\,\right)\,
\Gamma\left(\, 2- d_U\,\right)\,\left(\, p^2 \, \right)^{d_U-2}
\end{eqnarray}

 By using the relation between Euler gamma functions:

\begin{equation}
 z\,\Gamma\left(\, z\,\right)=\Gamma\left(\,1+ z\,\right)\ ,\quad
 \Gamma\left(\, z\,\right)\Gamma\left(\, -z\,\right)=
 -\frac{\pi}{z\,\sin\left(\, \pi\, z\,\right)}
\end{equation}

we recover the Euclidean form of the un-particle propagator, originally
found in \cite{Georgi:2007si}

 \begin{equation}
   G_U\left(\, p^2\,\right)=\frac{A_{d_U}}{2\,
\left(\,\Lambda^2_U\,\right)^{d_U-1}}\,
\frac{\left(\, p^2 \, \right)^{d_U-2}}{\sin\left(\, \pi\, d_U\,\right)}
\end{equation}

It is straightforward to guess the form of the corresponding
 \textit{effective action}:

\begin{equation}
S_\phi= \frac{\sin\left(\, \pi\, d_U\,\right)}{2A_{d_U}}\,  \int d^4x \,\,
\partial_\mu \phi\,\left(\, \frac{-\partial^2}{\Lambda^2_U}\,\right)^{1-d_U}
\partial^\mu \phi \label{ueff}
\end{equation}

As the exponent $1-d_U$ is a real number, a more explicit form of
(\ref{ueff}) can be obtained by introducing a Schwinger
parametrization for higher order D'Alembertian operator

\begin{equation}
S_\phi= \frac{\sin\left(\, \pi\, d_U\,\right)}{2A_{d_U}\,
\Gamma\left(\,d_U-1\,\right)}\,  \int d^4x \,\int_0^\infty ds\, s^{d_U-2}\,
\partial_\mu \phi\,\mathrm{tr}\left[\,
e^{-s\left(\,-\partial^2/\Lambda^2_U\,\right)}\,
\right]\,\partial^\mu \phi \label{useff}
\end{equation}

Furthermore, Eq.(\ref{pot2}) allows to determine the corresponding static
potential.

\begin{eqnarray}
V_U\left(\, \vec{x}\,\right)&&=
4\pi\,\kappa \,\int \frac{d^3q}{\left(\, 2\pi\,\right)^3}\,
e^{i\vec{q}\cdot\vec{x}}\,\frac{A_{d_U}}{2\,
\left(\,\Lambda^2_U\,\right)^{d_U-1}}\,
\frac{\left(\, \vec{q}^{\,2} \, \right)^{d_U-2}}{\sin\left(\, \pi\, d_U\,\right)}
\nonumber\\
&&=4\pi\,\kappa \,\frac{A_{d_U}}{
\Lambda^{2d_U-2}_U\, \sin\left(\, \pi\, d_U\,\right)}\,
\frac{1}{\pi^{3/2}2^{5-d_U}}\,
\frac{\Gamma\left(\, d_U-1/2\,\right)}{\Gamma\left(\, 2- d_U\,\right)}
\frac{1}{\vert\, \vec{x}\,\vert^{\, 2d_U-1}}
\end{eqnarray}

A slightly simpler form can be obtained by inserting the explicit form
$A_{d_U}$. Thus, we get


\begin{equation}
V_U\left(\, \vec{x}\,\right)= \frac{\kappa\,\Gamma\left(\,
d_U+1/2\,\right)\,\Gamma\left(\, d_U-1/2\,\right)}
{\Lambda^{2d_U-2}_U\left(\, 2\pi\,\right)^{2d_U-1}\Gamma\left(\,
2d_U\,\right)} \frac{1}{\vert\, \vec{x}\,\vert^{\, 2d_U-1}}
\label{unp1}
\end{equation}
An alternative way to compute the static potential will be discussed in the
next section. \\

Our computational method is \textit{apparently} limited to non-gauge
particles, as gauge invariance forbids the presence of a mass term
in the classical action. This is not strictly correct as it is known
from the early work by Stuckelberg \cite{Stuckelberg:1951gg} that
vector particles can be massive without spoiling gauge invariance
provided  compensating fields are properly introduced
\cite{Burnel:1986kb,Burnel:1986kc}. In modern language the extra
degree of freedom to be added into the Proca-Maxwell Lagrangian is
the Goldstone boson $\theta\left(\, x\,\right) $. The Stuckelberg
approach has been subsequently superseded by the Higgs mechanism, as
the proper way to provide gauge vector bosons mass, but it still
represents a viable alternative in all those cases where a symmetry
breaking potential is not available, e.g. in the framework of
relativistic extended objects
\cite{Aurilia:1993qi,Smailagic:1999qw,Ansoldi:2000qs,Smailagic:2000hr,Ansoldi:2001xi}.

The Proca-Maxwell gauge invariant Lagrangian can be written as

\begin{equation}
L= \frac{1}{4}\, F_{\mu\nu}\, F^{\mu\nu}+\frac{m^2}{2}\left(\,
A_\mu - \frac{1}{e}\,\partial_\mu\theta\,\right)^2
\end{equation}

where, $\theta$ is the Stuckelberg compensator which under gauge
transformation shifts as the Goldstone boson, that is

\begin{equation}
\theta\longrightarrow \theta + e\lambda
\end{equation}

The generating functional for the gauge field $A$ can be obtained by
integrating $\theta$ in the path integral

\begin{eqnarray}
 Z_A\left(\, J\,\right)&&=\int \left[\, DA\,\right]\,\left[\, D\theta\,\right]
 \,\exp\left(\, -\int d^4x\, L\,\right)\nonumber\\
 &&=\int \left[\, DA\,\right]\,\exp\left[\, -\int
 d^4x\, \left(\, -\frac{1}{4}\, F_{\mu\nu}\left(\, 1 - \frac{m^2}{\partial^2}\,
\right)\, F^{\mu\nu}
 + J^\mu\,A_\mu\,\right)\,\right]\nonumber\\
 &&
 \end{eqnarray}
Notice that, despite the presence of a mass term, the generating
functional is gauge invariant as the kinetic term is formulated
through the field strength only and the source $J$ is assumed to be
divergence-free. Thus, we get
\begin{eqnarray}
 Z_U\left(\, J\,\right)=&&\frac{A_{d_U}}{2\,
\left(\,\Lambda^2_U\,\right)^{d_U-1}}\,
\int_0^\infty dm^2
\, \left(\, m^2\,\right)^{d_U-2}\,
 \int \left[\, DA\,\right]\times\nonumber\\
 &&\exp\left[\, -\int
 d^4x\, \left(\, -\frac{1}{4}\, F_{\mu\nu}\left(\, 1 - \frac{m^2}{\partial^2}\,
\right)\, F^{\mu\nu}
 + J^\mu\,A_\mu\,\right)\,\right]
 \end{eqnarray}

Integration over the mass spectrum gives the un-photon Green function

\begin{equation}
G_U^{\mu\nu}\left(\,x\,\right)=\left(\, \delta^{\mu\nu}-
\frac{\partial^\mu\,\partial^\nu}{\partial^2}\,\right)\int \frac{d^4p}{\left(\,
2\pi\,\right)^4}\, e^{ipx}\, G_U\left(\, p^2\,\right)\label{ua}
\end{equation}

It is again straightforward to build an effective action leading to
the propagator (\ref{ua}):

\begin{equation}
S_A= -\frac{\sin\left(\, \pi\, d_U\,\right)}{4A_{d_U}}\,  \int d^4x \,\,
F_{\mu\nu}\,\left(\, \frac{-\partial^2}{\Lambda^2_U}\,\right)^{1-d_U}\,
F^{\mu\nu}\label{umax}
\end{equation}

\section{Alternative computation of the static potential}

As already mentioned, in this section we discuss an alternative
derivation of the static potential (\ref{unp1}), which is
distinguished by particular attention to gauge invariance. To do
this, we shall compute the expectation value of the energy operator
$H$ in the physical state $|\Phi\rangle$ describing the sources,
which we will denote by $ {\langle H\rangle}_\Phi$. Our starting
point is the Lagrangian density \cite{Krasnikov:2007fs}:
\begin{equation}
\mathcal{L} = \sum\limits_{k = 1}^N {\left[ { - \frac{1}{{4e_k^2 }}
F^{k\mu \nu } F_{\mu \nu }^k  + \frac{{m_k^2 }}{{2e_k^2 }}\left(
{A_\mu ^k  - \partial _\mu  \varphi ^{k} } \right)^2 } \right]},
\label{unp05}
\end{equation}
where $m_{k}$ is the mass for the $N$ scalar fields.

Following our earlier procedure \cite{GaeteSchSpa}, integrating out
the $\varphi$-fields induces an effective theory for the
$A_{\mu}^{k}$-fields. Once this is done, we arrive at the following
effective Lagrangian density:
\begin{equation}
\mathcal{L} = \sum\limits_{k = 1}^N {\frac{1}{{e_k^2 }}} \left[ { -
\frac{1}{4}F_{\mu \nu }^k \left( {1 + \frac{{m_k^2 }}{\Delta }}
\right)F^{k\mu \nu } } \right]. \label{unp10}
\end{equation}

Having characterized the theory under study, we can now compute the
interaction energy for a single mode in Eq. $(\ref{unp10})$. To this
end, we shall first examine the Hamiltonian framework for this
theory . The canonical momenta  $\Pi ^\mu   = - \left( {1 +
\frac{{m_{k}^2 }}{{\Delta}}} \right)F^{0\mu }$, which results in the
usual primary constraint, $\Pi_0=0$, and $\Pi ^i  = \left( {1 +
\frac{{m_{k}^2 }}{{\Delta
 }}} \right)F^{i0}$. The canonical Hamiltonian is then
\begin{equation}
H_C  = \int {d^3 } x\left\{ { - \frac{1}{2}\Pi ^i \left( {1 +
\frac{{m_{k}^2 }}{{\Delta }}} \right)^{ - 1} \Pi _i  + \Pi ^i
\partial _i A_0  + \frac{1}{4}F_{ij} \left( {1 + \frac{{m_{k}^2
}}{{\Delta }}} \right)F^{ij} } \right\}.\label{unp15}
\end{equation}
Time conservation of the primary constraint $ \Pi _0$ leads to the
secondary Gauss-law constraint $\Gamma _1 \left( x \right) \equiv
\partial _i
 \Pi ^i = 0$. The preservation of $\Gamma_1$ for all times does not
give rise to any further constraints. The theory is thus seen to
possess only two constraints, which are first class, therefore the
theory described by $(\ref{unp10})$ is a gauge-invariant one. The
extended Hamiltonian that generates translations in time then reads
$H = H_C  + \int {d^3 } x\left( {c_0 \left( x \right)\Pi _0 \left( x
\right) + c_1 \left( x \right)\Gamma _1 \left( x \right)} \right)$,
where $c_0 \left( x \right)$ and $c_1 \left( x \right)$ are the
Lagrange multiplier fields. Moreover, it is straightforward to see
that $\dot{A}_0 \left( x \right)= \left[ {A_0 \left( x \right),H}
\right] = c_0 \left( x \right)$, which is an arbitrary function.
Since $ \Pi^0 = 0$ always, neither $ A^0 $ nor $ \Pi^0 $ are of
interest in describing the system and may be discarded from the
theory. Then, the Hamiltonian takes the form
\begin{equation}
H = \int {d^3 x\left\{ { - \frac{1}{2}\Pi _i \left( {1 +
\frac{{m_{k}^2 }}{{\Delta}}} \right)^{ - 1} \Pi ^i  +
\frac{1}{4}F_{ij} \left( {1 + \frac{{m_{k}^2 }}{{\Delta}}}
\right)F^{ij}  + c\left( x \right){\partial _i} \Pi ^i} \right\}},
\label{unp20}
\end{equation}
where $c(x) = c_1 (x) - A_0 (x)$.

The quantization of the theory requires the removal of nonphysical
variables, which is done by imposing a gauge condition such that the
full set of constraints becomes second class. A convenient choice is
found to be \cite{GaeteZ}
\begin{equation}
\Gamma _2 \left( x \right) \equiv \int\limits_{C_{\xi x} } {dz^\nu }
A_\nu \left( z \right) \equiv \int\limits_0^1 {d\lambda x^i } A_i
\left( {\lambda x} \right) = 0, \label{unp25}
\end{equation}
where  $\lambda$ $(0\leq \lambda\leq1)$ is the parameter describing
the spacelike straight path $ x^i = \xi ^i  + \lambda \left( {x -
\xi } \right)^i $, and $ \xi $ is a fixed point (reference point).
There is no essential loss of generality if we restrict our
considerations to $ \xi ^i=0 $. In this case, the only non-vanishing
equal-time Dirac bracket is
\begin{equation}
\left\{ {A_i \left( x \right),\Pi ^j \left( y \right)} \right\}^ *
=\delta{ _i^j} \delta ^{\left( 3 \right)} \left( {x - y} \right) -
\partial _i^x \int\limits_0^1 {d\lambda x^j } \delta ^{\left( 3
\right)} \left( {\lambda x - y} \right). \label{unp30}
\end{equation}

We now turn to the problem of obtaining the interaction energy
between point-like sources in the model under consideration. As
mentioned above, we will work out the expectation value of the
energy operator $H$ in the physical state $\left| \Phi
\right\rangle$. Now we recall that the physical states $\left| \Phi
\right\rangle$ are gauge-invariant \cite{Dirac}. In that case we
consider the stringy gauge-invariant state
\begin{equation}
\left| \Phi  \right\rangle  \equiv \left| {\overline \Psi  \left(
\bf y \right)\Psi \left( {\bf y}\prime \right)} \right\rangle  =
\overline \psi \left( \bf y \right)\exp \left( {iq\int\limits_{{\bf
y}\prime}^{\bf y} {dz^i } A_i \left( z \right)} \right)\psi
\left({\bf y}\prime \right)\left| 0 \right\rangle, \label{unp35}
\end{equation}
where $\left| 0 \right\rangle$ is the physical vacuum state and the
line integral appearing in the above expression is along a
space-like path starting at ${\bf y}\prime$ and ending $\bf y$, on a
fixed time slice. It is worth noting here that the strings between
fermions have been introduced in order to have a gauge-invariant
function $\left| \Phi  \right\rangle $. In other terms, each of
these states represents a fermion-antifermion pair surrounded by a
cloud of gauge fields sufficient to maintain gauge invariance.

Next, from our above Hamiltonian analysis, we note that
\begin{equation}
\Pi _i \left( x \right)\left| {\overline \Psi  \left( {\bf y}
\right)\Psi \left( {\bf y^\prime} \right)} \right\rangle  =
\overline \Psi \left( {\bf y} \right)\Psi \left( {\bf y^\prime}
\right)\Pi _i \left( x \right)\left| 0 \right\rangle  +
q\int\limits_{\bf y}^{\bf y^\prime} {dz_i \delta ^{(3)} \left( {{\bf
z} - {\bf x}} \right)\left| \Phi \right\rangle }.\label{unp40}
\end{equation}
Having made this observation and since the fermions are taken to be
infinitely massive (static) we can substitute $\Delta$ by $- \nabla
^2$ in  Eq. $(\ref{unp20})$. In such a case $\left\langle H
\right\rangle _\Phi$ reduces to
\begin{equation}
\left\langle H \right\rangle _\Phi   = \left\langle H \right\rangle
_0  + V, \label{unp45}
\end{equation}

where $\left\langle H \right\rangle _0  = \left\langle 0
\right|H\left| 0 \right\rangle$. The $V$ term is given by:
\begin{equation}
V =  - \frac{{q^2 }}{2}\int {d^3 x} \int_{\bf y}^{\bf y^\prime}
{dz^\prime_i } \delta ^{\left( 3 \right)} \left( {x - z^\prime}
\right)\frac{1}{{\nabla _x^2  - m_{k}^2 }}\nabla _x^2 \int_{\bf
y}^{\bf y^\prime} {dz^i } \delta ^{\left( 3 \right)} \left( {x - z}
\right), \label{unp50}
\end{equation}
where the integrals over $z^i$ and $z^\prime_i$ are zero except on
the contour of integration. This term may look peculiar, but it is
just the familiar Yukawa interaction plus self-energy terms. In
effect, expression (\ref{unp50}) can also be written as
\begin{equation}
V = \frac{{q^2 }}{2}\int_{\bf y}^{{\bf y}^{\prime}  }
{dz_i^{\prime}}\partial _i^{z^{\prime}} \int_{\bf y}^{{\bf
y}^{\prime}} {dz^i }\partial _z^i G\left( {{\bf z}^{\prime},{\bf z}}
\right), \label{unp55}
\end{equation}
where $G$ is the Green function
\begin{equation}
G({\bf z}^{\prime}  ,{\bf z}) = \frac{1}{{4\pi }}\frac{{-e^{ {m_k}
{\left| {{\bf z}^ {\prime}   - {\bf z}} \right|} }}}{{\left| {{\bf
z}^ {\prime}   - {\bf z}} \right|}}. \label{unp60}
\end{equation}
Employing Eq.(\ref{unp60}) and remembering that the integrals over
$z^i$ and $z_i^{\prime}$ are zero except on the contour of
integration, expression (\ref{unp55}) reduces to the familiar Yukawa
interaction after subtracting the self-energy terms. Therefore the
potential for two opposite charges located at ${\bf y}$ and ${\bf
{y^\prime}}$ is given by
\begin{equation}
V = - \frac{{q^2 }}{{4\pi }}\frac{{e^{ {-m_k } L} }}{L}.
\label{unp65}
\end{equation}
where $L\equiv|{\bf y}-{\bf {y^\prime}}|$. However, from
(\ref{unp10}) we must sum over all the modes in (\ref{unp60}), that
is,
\begin{equation}
V =  - \frac{{q^2 }}{{4\pi }}\sum\limits_{k = 1}^N {\frac{1}{{e_k^2
}}} \frac{{e^{ - m_k L} }}{L}. \label{unp66}
\end{equation}
In effect, as was explained in Ref. \cite{Krasnikov:2007fs}, in the
limit $k
 \to\infty$ the sum is substituted by an integral as follows
\begin{equation}
V = \left( { - \frac{{q^2 }}{{4\pi }}} \right)\frac{1}{L}A_{d_U }
\int\limits_0^\infty  {t^{d_U  - 2} } e^{ - \sqrt t L} dt,
\label{unp70}
\end{equation}
where $t=m_{k}^{2}$. Here ${t^{d_U  - 2} }$ is the spectral density,
and $A_{d_{U}}$ is a normalization factor which is given by
expression (\ref{uz2}). A direct computation on the $t$-variable
yields
\begin{equation}
V = \left( { - \frac{{q^2 }}{{4\pi }}} \right)A_{d_U }
\frac{2}{{L^{2d_U  - 1} }}\Gamma \left( {2d_U  - 2} \right).
\label{unp75}
\end{equation}
Using the relationship involving Gamma functions, that is,
\begin{equation}
\Gamma \left( {2d_u  - 2} \right) = \left( {2\pi } \right)^{ - {1
\mathord{\left/
 {\vphantom {1 2}} \right.
 \kern-\nulldelimiterspace} 2}} 2^{2d_u  - {5 \mathord{\left/
 {\vphantom {5 2}} \right.
 \kern-\nulldelimiterspace} 2}} \Gamma \left( {d_U  - 1} \right)\Gamma
 \left( {d_U  - {\raise0.7ex\hbox{$1$} \!\mathord{\left/
 {\vphantom {1 2}}\right.\kern-\nulldelimiterspace}
\!\lower0.7ex\hbox{$2$}}} \right), \label{unp80}
\end{equation}
expression (\ref{unp75}) then becomes
\begin{equation}
V = \left( { - \frac{{q^2 }}{{4\pi }}} \right)\frac{4}{{\pi ^{2d_U
- 1} }} \frac{{\Gamma \left( {d_U  + {\raise0.7ex\hbox{$1$}
\!\mathord{\left/
 {\vphantom {1 2}}\right.\kern-\nulldelimiterspace}
\!\lower0.7ex\hbox{$2$}}} \right)\Gamma \left( {d_U  -
{\raise0.7ex\hbox{$1$} \!\mathord{\left/
 {\vphantom {1 2}}\right.\kern-\nulldelimiterspace}
\!\lower0.7ex\hbox{$2$}}} \right)}}{{\Gamma \left( {2d_U } \right)}}
\frac{1}{{\left( L \right)^{2d_U  - 1} }}. \label{unp85}
\end{equation}
By introducing the scale factor $ l = \frac{1}{{2^{\frac{1}{{2d_U  -
2}}} \Lambda _U }}$, the corresponding modified Coulomb potential
may be written as
\begin{equation}
V = \left( { - \frac{{q^2 }}{{4\pi }}} \right) \frac{1}{L} \left[ {1
+ \frac{2}{{\pi ^{2d_U  - 1} }}\frac{{\Gamma \left( {d_U  +
{\raise0.7ex\hbox{$1$} \!\mathord{\left/
 {\vphantom {1 2}}\right.\kern-\nulldelimiterspace}
\!\lower0.7ex\hbox{$2$}}} \right)\Gamma \left( {d_U  -
{\raise0.7ex\hbox{$1$} \!\mathord{\left/
 {\vphantom {1 2}}\right.\kern-\nulldelimiterspace}
\!\lower0.7ex\hbox{$2$}}} \right)}}{{\Gamma \left( {2d_U }
\right)}}\left( {\frac{l}{L}} \right)^{2d_U  - 2} } \right].
\label{unp90}
\end{equation}

The above potential profile is analogous to the one encountered in
\cite{Goldberg:2007tt}.

\section{Concluding Remarks}

Having obtained (\ref{umax}) it is tempting to extend this result both to
Yang-Mills fields and to gravity.\\
In the first case, the passage from Abelian to non-Abelian un-vector
gauge boson start from the massive Yang-Mills non-local action,
which can be formally written as

\begin{equation}
S^A_U= -\frac{1}{4}\,  \int d^4x \,\,\mathrm{tr}\,\left[\,
\mathbf{F}_{\mu\nu}\,\left(\, 1+ m^2\, \left(\,
-\mathcal{D}_\mu\,\mathcal{D}^\mu\, \right)^{-1}\,\right)\,\mathbf{F}^{\mu\nu}
\,\right]\label{ymm}
\end{equation}

where, the inverse gauge D' Alembertian operator can be defined through a
perturbative series in the coupling constant.\\
Integration over $m$ leads to the non-Abelian version of
(\ref{umax})

\begin{equation}
S_U^{YM}= -\frac{\sin\left(\, \pi\, d_U\,\right)}{4A_{d_U}}\,  \int d^4x \,\,
\mathrm{tr}\,\left[\,
\mathbf{F}_{\mu\nu}\,\left(\, \frac{-\partial^2}{\Lambda^2_U}\,\right)^{1-d_U}
\,\mathbf {F}^{\mu\nu}\,\right]\label{uym}
\end{equation}

By choosing a static charge distribution for $J^\mu$, one recovers the
un-particle correction to the Coulomb potential \cite{Goldberg:2007tt}.\\
Colored un-particle dynamics has been perturbatively investigated in
\cite{Cacciapaglia:2007jq}, where the production cross section for scalar
unparticles mediated by a gauge interaction has been computed. More in detail,
Eq.(3.1) in \cite{Cacciapaglia:2007jq} is the momentum space version
of Eq.(\ref{ueff}) with the addition of an infrared cut-off $m$, but the
gauge effective action Eq.(4.1) is simply the ordinary effective action
times a factor $(2-d_U)$.  As a consequence the production cross section
at the first order in the coupling constant
results to be $(2-d_U)$ times the standard result.\\
On the other hand, our effective action (\ref{uym}), is a gauge invariant
\textit{non-perturbative} result taking into account un-particle effects
to any order in the coupling constant. To our knowledge, equations
(\ref{umax}), (\ref{uym}) are new results.
\\
The un-gravity effective action is little more involved as one
starts from linearized gravity. Un-particle modifications to the Newton
potential has also been discussed in \cite{Goldberg:2007tt} and possible
test in planetary system in \cite{Das:2007cc}.
The action for linearized, general coordinate transformation respecting,
massive gravity can be written as

\begin{equation}
S^h= \frac{1}{2}\,  \int d^4x \,
h_{\mu\nu}\,\left(\, 1+ \frac{m^2}{-\partial^2}\,\right)\,
\Delta^{\mu\nu\alpha\beta} \, h_{\alpha\beta}
\label{ugraviton}
\end{equation}

where, $\Delta^{\mu\nu\alpha\beta}$ is the covariant D'Alembertian
for rank-two symmetric tensors.\\
Once more, integration over the mass spectrum leads to the
un-graviton effective action

\begin{equation}
S_U^h= \frac{\sin\left(\, \pi\, d_U\,\right) }{2A_{d_U} }\,  \int d^4x \,
h_{\mu\nu}\, \,
\Delta^{\mu\nu\alpha\beta} \,
\left(\, \frac{-\partial^2}{\Lambda^2_U}\,\right)^{1-d_U}\, h_{\alpha\beta}
\end{equation}

Now, it is tempting to conjecture the exact form of
gravity action in the un-particle sector. As the graviton kinetic
term comes from the weak field expansion of the Ricci scalar, $R$
,and the effect of integrating over the mass spectrum amounts to
modify the standard action by inserting the operator
$\left(\,-\partial^2/\Lambda^2_U\,\right)^{1-d_U}$, we conclude that

\begin{equation}
S_U^h= \frac{\sin\left(\, \pi\, d_U\,\right) }{32\pi\, G_N\,
A_{d_U} }\,\int d^4x \sqrt{g}\,
\left(\, \frac{-\Delta^\mu\,\partial_\mu}{\Lambda^2_U}\,\right)^{1-d_U}R
\label{ueh}
\end{equation}

where, $\Delta_\mu$ is the generally covariant derivative. To prove that
this is the correct effective action reproducing (\ref{ugraviton}) one has
to remember that the covariant derivative is metric compatible, thus the
covariant D'Alembertian acting on the Ricci, $R=g^{\mu\nu}\, R_{\mu\nu}$,
scalar is blind to $g^{\mu\nu}$ and can freely moved in front of the Ricci
tensor.\\
To conclude, we summarize the main results of this paper.
We have constructed effective actions for scalar,
gauge vector and tensor un-particles by starting from the explicit
expression of the Feynman propagator.
While the scalar effective was
already available in the literature \cite{Krasnikov:2007fs}, the
Abelian/Nonabelian and gravitational cases are new results obtained
through a proper implementation of the Stuckelberg compensating
method. In alternative to the known
procedure, based on conformal scaling arguments, we explicitly
evaluated the Green function through standard functional techniques
supplemented by a mass spectrum integration. In the case of gauge
un-particle we applied the mass integration to a Stuckelberg type
action for massive gauge vectors. The gauge invariant effective
action (\ref{umax}) has been extended to the Yang-Mills case
(\ref{uym}). The same procedure has been applied to tensor
un-particles to recover the gravitational effective action
(\ref{ueh}).  All these results are new ones and deserve
further investigations.\\
Finally, we recovered the static potential generated by un-particle
exchange in a gauge-invariant path dependent framework, which has
been previously introduced to study color confinement in Yang-Mills
theories. An important feature of this methodology is that it
provides a physically-based alternative to the usual Wilson loop
approach.

\section{ACKNOWLEDGMENTS}
P. G. was partially supported by Fondecyt (Chile) grant 1050546.

\end{document}